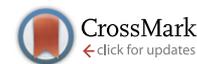

# One-step architecture of bifunctional petal-like oxygen-deficient NiAl-LDHs nanosheets for high-performance hybrid supercapacitors and urea oxidation

Yuchen Wang, Yaoyu Liu, Man Zhang, Biying Liu, Zhiyue Zhao and Kai Yan*

**ABSTRACT** Nickel-based layered double hydroxides (LDHs) are promising electrode materials in the fields of energy storage (supercapacitors) and conversion (urea oxidation). The rational construction of atomic and electronic structure is crucial for nickel-based LDHs to realize their satisfactory electrochemical performance. Herein, we report a facile, eco-friendly, one-step synthesis process to construct petal-like oxygen-deficient NiAl-LDH nanosheets for hybrid supercapacitors (HSCs) and urea oxidation reaction (UOR). The as-prepared NiAl-LDH nanosheets with rich oxygen vacancies possess a large specific surface area of 216.6 m$^2$ g$^{-1}$ and a desirable electronic conductivity of $3.45 \times 10^{-4}$ S cm$^{-1}$ to deliver an ultra-high specific capacitance of 2801 F g$^{-1}$ (700 C g$^{-1}$) at 1 A g$^{-1}$. Furthermore, high specific energy of 50.0 W h kg$^{-1}$ at 400 W kg$^{-1}$ and excellent cycle stability with 91% capacitance retention after 10,000 cycles are achieved by the NiAl-LDHs/CFP (carbon fiber paper) (+)//YP-80F (a commercial activated carbon) (−) HSC. Besides, NiAl-LDH nanosheets also work as an efficient electrocatalyst for UOR, which only requires 1.42 V *vs.* reversible hydrogen electrode to drive 10 mA cm$^{-2}$ in 1 mol L$^{-1}$ KOH with 0.33 mol L$^{-1}$ urea. This remarkable performance is superior to most reported values of previous candidates owing to the thin structure of NiAl-LDH nanosheets for exposing more active sites and abundant oxygen vacancies. In addition, various reaction parameters are investigated to optimize the electrochemical performance. In general, this work paves a new way for the architecture of multifunctional nanostructured energy materials.

**Keywords:** layered double hydroxides, nanosheets, oxygen vacancy, hybrid supercapacitor, urea oxidation reaction

## INTRODUCTION

The global trend of carbon neutrality promotes the development of novel electrochemical energy storage and conversion (ESC) technologies to alleviate the use of fossil energy sources [1–3]. Among various emerging technologies, hybrid supercapacitors (HSCs) and urea electrolysis for hydrogen are considered as efficient representatives to overcome the energy crisis. Specifically, HSCs are constructed to produce high energy and power simultaneously *via* a combined non-Faradaic capacitive and Faradaic battery-like energy storage process [4,5]. However, compared with current cutting-edge battery technologies, one major bottleneck of HSCs is the relatively low specific energy, which is mainly determined by battery-like positive electrode materials [6]. On the other hand, urea electrolysis is expected to replace the conventional water splitting for producing hydrogen energy in alkaline media, owing to its prominently lower theoretical potential at the anode side (−0.46 V *vs.* standard hydrogen electrode (SHE)) in comparison with that of water splitting (0.40 V *vs.* SHE) [7]. However, the complex anodic urea oxidation reaction (UOR) involving six-electron transfer leads to the low kinetics and hampers the overall electrocatalytic efficiency of urea electrolysis [8–10]. Thus rational design of low-cost and high-performance bifunctional electrode materials is urgently needed for both advanced ESC systems.

Nickel-based layered double hydroxides (LDHs) have drawn tremendous attention in various applications including HSCs, oxygen evolution reaction (OER), and UOR due to their unique lamellar structure and tunable transition metal compositions [11–13]. In comparison with other types of LDHs materials, the mechanism of nickel-based LDHs for both HSCs and UOR involves the same redox reaction of Ni(OH)$_2$ + OH$^-$ ↔ NiOOH + H$_2$O + e$^-$ [14–16]. To achieve optimized electrochemical performance of these nickel-based LDHs, numerous efforts have been conducted to tune their atomic and electronic structure ($t_{2g}^6 e_g^0 / t_{2g}^6 e_g^1$), including morphology regulation [17,18], defect engineering [19], and complexation with highly conductive materials [20]. For instance, Tang *et al.* [21] synthesized flower-like oxygen-vacancy abundant NiMn-LDH nanosheets with a maximum specific capacity of 1183 C g$^{-1}$ *via* electrodeposition and subsequent H$_2$O$_2$ oxidation. Sun *et al.* [22] developed Rh-doped NiFe-LDH nanosheets with rich oxygen vacancies through an *in-situ* growth method with the aid of ethylene glycol. These NiFe-LDH nanosheets exhibit high UOR activity to deliver a current density of 10 mA cm$^{-2}$ with a potential of 1.35 V *vs.* reversible hydrogen electrode (RHE). Therefore, the oxygen-deficient nanosheet structure of nickel-based LDHs is essential for realizing their outstanding supercapacitive and UOR performance. The nanosheet structure is favorable for the exposure of more active sites and the rapid accessibility of electrolyte ions [23]. Additionally, the existence of abundant oxygen vacancies is conducive to enhancing electronic conductivity and decreasing absorption energy of OH$^-$ [21]. However, up to present, no work has reported a facile and universal strategy to develop bifunctional oxygen-deficient nickel-based LDH nanosheets with high-performance for both ESC systems.

Herein, we propose a facile one-step methodology to *in situ* grow oxygen-deficient NiAl-LDH nanosheets on carbon fiber paper (CFP). The fabricated NiAl-LDHs/CFP electrodes exhibit superior supercapacitive performance and high electrocatalytic activity for UOR. The novel electrodes are constructed with the following features: (1) the eco-friendly *in situ* growth process of NiAl-LDH nanosheets does not involve any organic solvents and alkaline sources (e.g., urea, sodium hydroxide); (2) the architecture of petal-like NiAl-LDH nanosheets with abundant oxygen vacancies could provide plenty of active sites, promote the intrinsic conductivity, and facilitate the interactions between LDHs and electrolyte ions; (3) the as-obtained NiAl-LDHs/CFP electrodes are free from binder or external conductive chemicals. This work not only provides a new synthesis approach to prepare oxygen-vacancy-abundant LDH nanostructured materials, but also highlights a path for their potential in realizing dual functions as HSC electrode materials and UOR electrocatalysts.

## EXPERIMENTAL SECTION

### Hydrophilic modification of CFP

Commercial CFP (Spectracarb 2050A-1050, Fuel Cell Store) with a thickness of 254 μm was cut into 2 cm × 1 cm rectangles and ultrasonically cleaned in ultrapure water (18.25 MΩ cm), acetone (analytical reagent (AR), ≥99.5%, Guangzhou Chemical Reagent Factory), and ethanol (AR, ≥99.7%, GHTECH). Each ultrasonic treatment time was 0.5 h. Then the CFP was immersed in 5 mL of concentrated nitric acid (GR, 65.0%–68.0%, Guangzhou Chemical Reagent Factory) for 12 h. After surface modification, the CFP was washed with ethanol and ultrapure water several times to remove nitric acid.

### Preparation of NiAl-LDHs/CFP electrodes

One-step construction of NiAl-LDHs/CFP electrodes is depicted in Fig. 1a. First, 2.25 mmol of nickel nitrate hexahydrate ($Ni(NO_3)_2 \cdot 6H_2O$, guaranteed reagent, 99%, Aladdin) and 0.75 mmol of aluminum nitrate nonahydrate ($Al(NO_3)_3 \cdot 9H_2O$, 99.99% metals basis, Aladdin) powders were dissolved in 40 mL of an ultrapure water/ethanol mixture ($v/v = 1:1$). Subsequently, two pieces of modified CFP were placed at the bottom of a 100-mL Teflon reaction vessel, and the above solution was slowly dropped into the vessel. After hydrothermal treatment at 160°C for 8 h, NiAl-LDHs/CFP electrodes were collected by washing with ultrapure water and vacuum drying at 80°C for 1 h. The loading mass of active materials was 1.0–2.0 mg cm$^{-2}$. NiAl-LDH powders were obtained by centrifugation and drying under vacuum at 80°C for 8 h. NiAl-LDHs/CFP electrodes and NiAl-LDH powders prepared in this work were denoted as NAC-Ov and NA-Ov, respectively.

For comparison, NiAl-LDHs/CFP electrodes and NiAl-LDH powders were also prepared using a conventional hydrothermal method, which is described in the Supplementary information. The corresponding NiAl-LDHs/CFP electrodes and NiAl-LDH powders were denoted as NAC-U and NA-U, respectively. In addition, the hydrothermal treatment time ($x = 4, 8, 12$ h) was adjusted to determine the optimal conditions. NiAl-LDHs/CFP electrodes and NiAl-LDH powders with different hydrothermal treatment times were labeled as NAC-Ov-$x$ and NA-Ov-$x$, respectively.

The characterizations of material, supercapacitive, and electrocatalytic properties are depicted in the Supplementary information.

## RESULTS AND DISCUSSION

### Microstructure analysis

The crystal structures of NA-U and NA-Ov were first investigated by X-ray diffraction (XRD), as shown in Fig. 1b. For NA-U, the typical diffraction peaks at 11.4°, 23.0°, 34.9°, 39.3°, 46.6°, 60.9°, and 62.2° can be ascribed to the crystal planes of (003), (006), (012), (015), (018), (110), and (113) of NiAl-LDHs (PDF#15-0087), respectively [24], manifesting the successful construction of hydrotalcite-like structure. For NA-Ov, the diffraction peaks become relatively weak and broad, indicating a low degree of crystallization. It is also noteworthy that the diffraction peaks of (003) and (006) shift to low angles at 9.9° and 19.9°, respectively. As seen in Fig. 1c, the basal spacing of NA-Ov is calculated as 2.69 nm based on Bragg's law, which is larger than that of NA-U (2.33 nm). The interlayer spacing difference is mainly attributed to the anionic species between the layers [25]. In the synthesis process of NA-U, abundant $CO_3^{2-}$ ions are formed from the hydrolysis of urea and intercalated into interlayers. In contrast, $NO_3^-$ ions are dominant in the interlayers of NA-Ov, enlarging the interlayer spacing, which is beneficial for rapid diffusion of electrolyte ions and thus fast electrochemical kinetics. Additionally, the conductivity of NA-U and NA-Ov were measured using a four-point probe method at room temperature. As displayed in Table S1, the intrinsic conductivity of NA-Ov is around three times higher than that of NA-U, demonstrating the modification of electronic structure possibly induced by oxygen vacancies [26].

The Brunauer-Emmett-Teller (BET) specific surface area analysis is vital for determining the supercapacitive and electrocatalytic performance of NiAl-LDHs. The typical type IV isotherms of NiAl-LDHs (Fig. S1a) with obvious hysteresis loops demonstrate the mesoporous structure of both materials [27], which are confirmed by the pore size distribution with the range from 3 to 5 nm (Fig. S1b). The specific surface area of NA-Ov is calculated as 216.6 m$^2$ g$^{-1}$, which is larger than that of NA-U (186.9 m$^2$ g$^{-1}$). The large specific surface area is desirable for exposing more active sites to realize high charge storage and urea oxidation capabilities.

Morphologies and microstructure of NiAl-LDH nanosheets were then studied by atomic force microscopy (AFM), field emission scanning electron microscopy (FE-SEM), and transmission electron microscopy (TEM). Three dimensional (3D) views of the as-synthesized NiAl-LDHs in Fig. 1d, e distinctly show that NA-U and NA-Ov are composed of nanoflakes and nanosheets, respectively. According to the height profiles in Fig. S2, the corresponding thicknesses for NA-U and NA-Ov are estimated to be 3.8 and 2.3 nm, respectively. SEM images of NAC-Ov are illustrated in Fig. 2a, b. Petal-like nanosheets are uniformly formed on the surface of CFP. The well-defined petal-like nanostructure is advantageous for exposing active sites and promoting the penetration of electrolyte ions. However, for NAC-U, few nanoflakes are grown on the CFP substrate (Fig. 2c, d). TEM images of NA-Ov and NA-U (Fig. 2e, f) display similar morphologies with the corresponding AFM images. Intact circular nanoflakes are clearly observed for NA-U, whereas fragmented nanosheets with relatively small sizes are

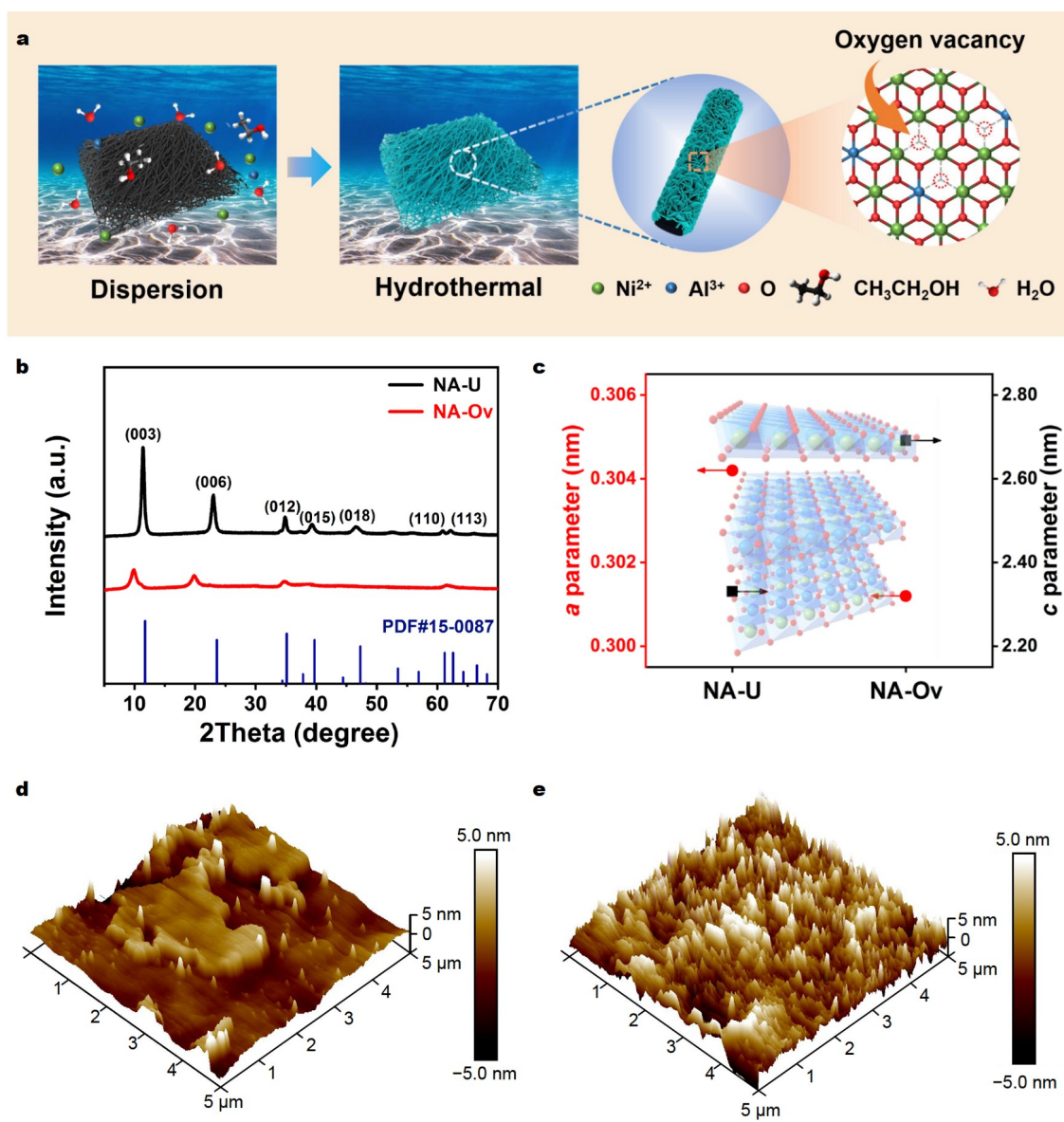

**Figure 1** (a) One-step synthesis of oxygen-deficient NiAl-LDHs/CFP electrodes. (b) XRD patterns, (c) lattice parameters of NA-U and NA-Ov. AFM 3D views of (d) NA-U and (e) NA-Ov.

captured for NA-Ov. Moreover, the crystal structure of NiAl-LDHs was further confirmed by high-resolution TEM (HR-TEM). The HR-TEM image and the corresponding selected area electron diffraction (SAED) pattern of NA-Ov (Fig. 2g, h) display an interplanar spacing of 0.197 nm which matches well with the crystal plane (018) of standard LDHs. Similarly, crystal planes (110) of NA-U with an interplanar spacing of 0.152 nm are indicated in Fig. S3. Additionally, energy dispersive X-ray spectroscopy (EDX) elemental mappings of NAC-Ov illustrate the homogeneous distribution of Ni, Al, and O elements on CFP (Fig. 2i).

Chemical valence states of elements and the concentration of oxygen vacancies were determined by X-ray photoelectron spectroscopy (XPS) and electron paramagnetic resonance (EPR). In Fig. 3a, the surface $Ni^{2+}$ and $Ni^{3+}$ in Ni $2p_{3/2}$ spectra are related to peaks at 855.6 and 856.8 eV, respectively [28]. According to Table S2, $Ni^{2+}/Ni^{3+}$ ratios of NA-U and NA-Ov are calculated as 0.87 and 1.55, respectively. The reduction of nickel chemical valence states reveals the formation of ample oxygen vacancies. Meanwhile, Al 2p peaks at 74.1 eV in Fig. 3b clarify the trivalent oxidation state of Al element in both NiAl-LDHs. In Fig. 3c, O 1s spectra are fitted into four peaks at 530.5, 531.5, 531.6, and 532.8 eV, which correspond to lattice oxygen ($O_L$), surface hydroxyl ($O_{OH}$), oxygen vacancy ($O_V$), and absorbed water ($O_W$), respectively [29]. Compared with NA-U, the larger peak area of $O_V$ in NA-Ov indicates the presence of a growing number of oxygen vacancies. Moreover, the stronger EPR signal of NA-Ov with $g = 1.999$ (Fig. 3d) further confirms that NA-Ov possesses a higher oxygen-vacancy concentration [30], which agrees well with the XPS results.

Combining the above XRD, conductivity, EPR, and XPS results, it is concluded that NiAl-LDHs nanosheets with abundant oxygen vacancies are successfully constructed by this facile one-step hydrothermal method. Oxygen-deficient NA-Ov pro-vides a large specific surface area (216.6 m$^2$ g$^{-1}$) and a high conductivity (3.45 × 10$^{-4}$ S cm$^{-1}$), which are key factors for achieving a superior electrochemical performance in both HSC and urea oxidation.

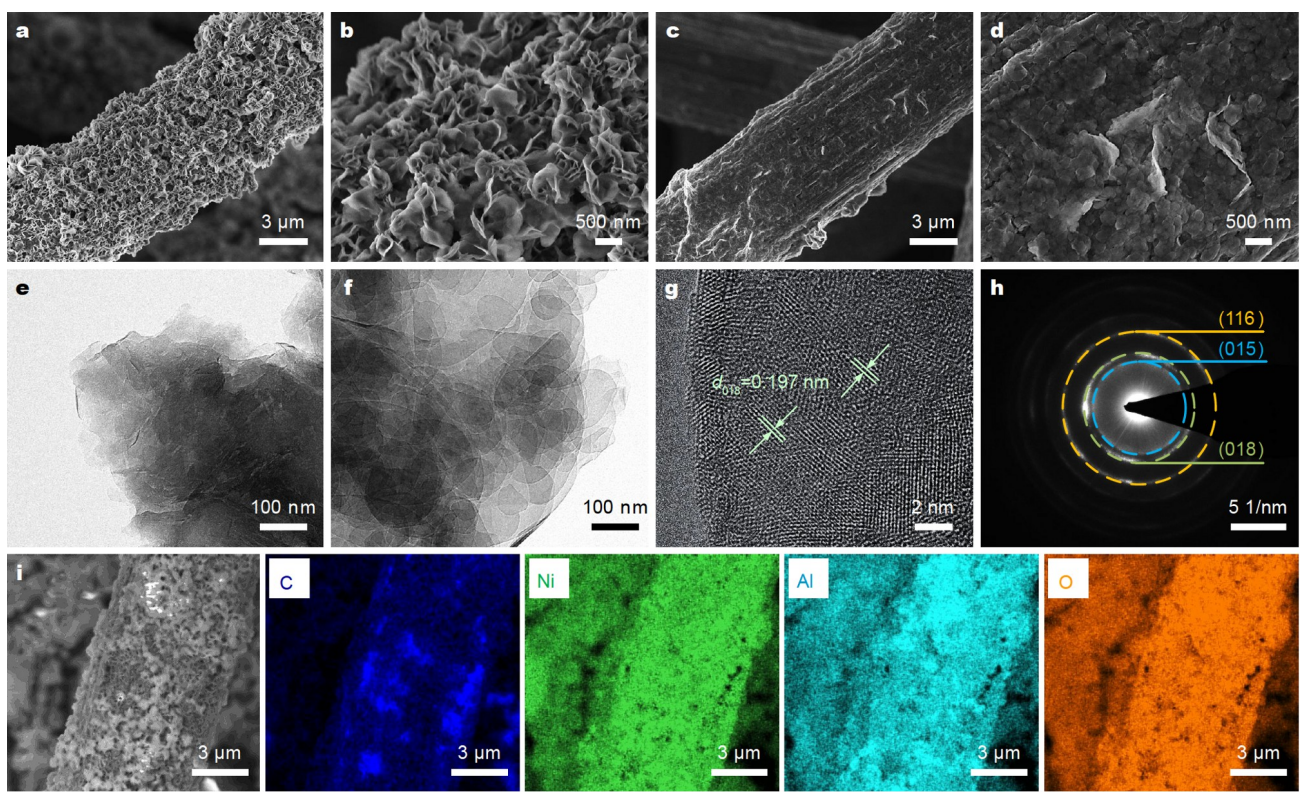

**Figure 2** FE-SEM images of (a, b) NAC-Ov and (c, d) NAC-U. TEM images of (e) NA-Ov and (f) NA-U. (g) HR-TEM and (h) the corresponding SAED images of NA-Ov. (i) EDX elemental mappings of NAC-Ov.

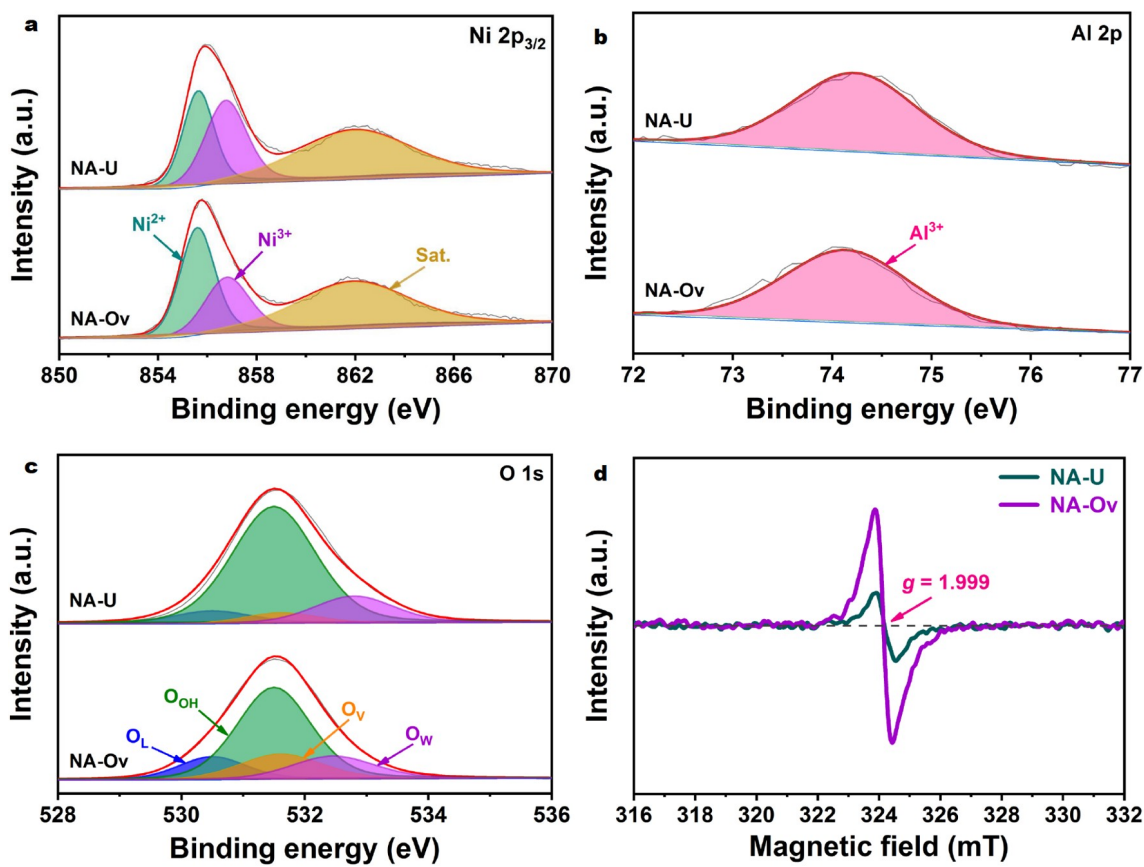

**Figure 3** XPS (a) Ni 2p spectra, (b) Al 2p spectra, (c) O 1s spectra and (d) EPR spectra of NA-U and NA-Ov.

**Supercapacitive behavior of NiAl-LDHs/CFP electrodes**

The supercapacitive behavior of the NiAl-LDHs/CFP electrodes was investigated by cyclic voltammetry (CV), galvanostatic charge/discharge (GCD), and electrochemical impedance spectroscopy (EIS) measurements in a three-electrode system. Fig. 4a displays the CV curves of NAC-U and NAC-Ov at a scan rate of 10 mV s$^{-1}$. A pair of well-defined symmetrical peaks in CV curves are mainly attributed to the Faradic redox reactions of Ni-O/Ni-O-OC (C represent K or H) [31]. The integrated area for NAC-Ov is clearly greater than that of NAC-U, demonstrating an enhanced charge storage ability. Fig. 4b presents a series of CV curves of NAC-Ov from 1 to 50 mV s$^{-1}$. At high scan rates, the potential difference between anodic and cathodic peaks becomes large due to the ohmic polarization and limited diffusion rate of electrolyte ions [32]. According to these CV curves, the corresponding charge storage behavior was analyzed using the power law between the peak current ($i$) and the scan rate ($v$): $i = av^b$ [33]. The $b$ values of 0.5 and 1.0 indicate a diffusion-controlled Faradaic process and a pseudocapacitive process, respectively. After fitting in Fig. 4c, the $b$ values are derived as 0.52 for NAC-U and 0.55 for NAC-Ov, revealing that the charge storage mechanism of both NiAl-LDHs/CFP electrodes is dominated by diffusion-controlled reactions.

As shown in Fig. 4d and Fig. S4, GCD curves of both electrodes display battery-like characteristics, which are in agreement with the analysis of CV curves. Based on these GCD curves, the corresponding specific capacitance values are summarized in Fig. 4e. Owing to the relatively large specific surface area and high-concentration oxygen vacancies, the maximum specific capacitance value of the NAC-Ov is 1.8 times higher than that of NAC-U. The NAC-Ov exhibits extremely high specific capacitance values of 2801, 2677, 2480, 2345, 2247, and 2160 F g$^{-1}$ (700, 669, 620, 586, 562, and 540 C g$^{-1}$) at specific currents of 1, 2, 4, 6, 8, and 10 A g$^{-1}$, respectively. Afterwards, Nyquist plots are exhibited in Fig. S5 to further study the beneficial effect of oxygen vacancy on electrochemical performance. By fitting the data with a classical equivalent circuit, impedance values including the equivalent series resistance $R_{ESR}$ and the charge transfer resistance $R_{ct}$ are obtained and listed in Table S3. Compared with NAC-U, the relatively low $R_{ESR}$ (0.9 Ω) and $R_{ct}$ (2.1 Ω) of NAC-Ov indicate that oxygen vacancies are conducive to promoting the intrinsic conductivity and the interaction between LDHs and electrolyte ions [21], which are consistent with the CV and GCD results.

To identify the material with the best supercapacitive performance, the hydrothermal time was adjusted from 4 to 12 h and the corresponding samples were investigated. From Fig. S6, the specific capacitance values of NAC-Ov-4, NAC-Ov-8, and NAC-Ov-12 are 1962 F g$^{-1}$ (491 C g$^{-1}$), 2801 F g$^{-1}$ (700 C g$^{-1}$), and 2403 F g$^{-1}$ (601 C g$^{-1}$) at 1 A g$^{-1}$, respectively. According to XRD results and porosity parameters (Figs S7, S8, and Table S4), the undesirable specific capacitance values of NAC-Ov-4 and NAC-Ov-12 are ascribed to the growth of Ni(OH)$_2$ phase (PDF#38-0715), the relatively low specific surface area and small pore size. Therefore, 8 h is chosen as the optimal hydrothermal time for the synthesis process and NAC-Ov-8 is renamed to NAC-Ov for simplification. To the best of our knowledge, the maximum specific value of NAC-Ov is superior to most reported values of NiAl-LDHs in relevant publications (Table S5). Next, the cycling performance of NAC-Ov was evaluated at a specific current of 10 A g$^{-1}$ (Fig. 4f). The specific capacitance remains 80% after 4000 cycles, indicating an excellent cycling stability of NAC-Ov.

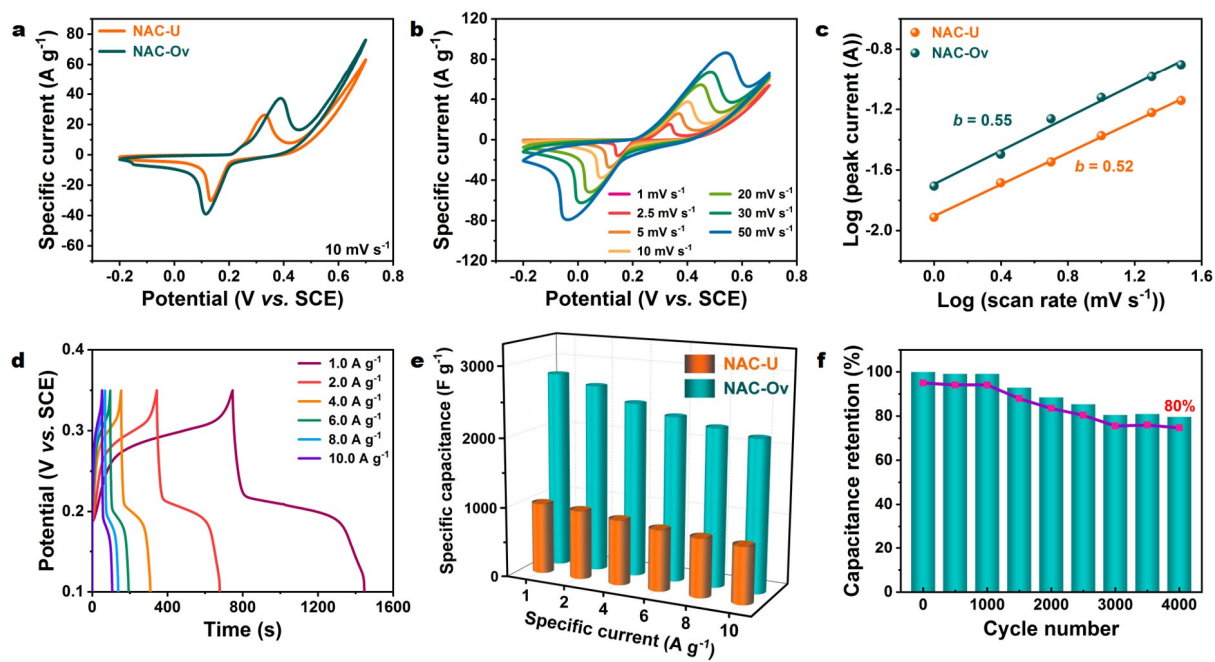

**Figure 4** (a) CV curves of NAC-U and NAC-Ov at a scan rate of 10 mV s$^{-1}$. (b) CV curves of NAC-Ov at scan rates of 1, 2.5, 5, 10, 20, 30, and 50 mV s$^{-1}$. (c) The $b$-value calculation with the reduction peak current and the scan rate for NAC-U and NAC-Ov. (d) GCD curves of NAC-Ov at specific currents of 1, 2, 4, 6, 8, and 10 A g$^{-1}$. (e) Specific capacitance values of NAC-Ov and NAC-U as a function of specific currents. (f) Cycle stability of NAC-Ov at 10 A g$^{-1}$.

To demonstrate the practical utilization of NAC-Ov, an HSC was fabricated with the configuration of NAC-Ov(+)//YP-80F (a commercial activated carbon) (−). Fig. S9 shows the CV curves of the positive electrode and the negative electrode at a scan rate of 10 mV s$^{-1}$ with the corresponding potential range from −0.2 to 0.7 V and from −0.9 to −0.2 V, respectively. Combined with Fig. S10, the optimal voltage window of the HSC is determined as 1.6 V. When the voltage is higher than 1.6 V, the current response rises sharply owing to the superfluous side reaction of water splitting [34]. Fig. 5a shows the CV curves of the HSC at various scan rates with broad redox peaks, implying a pseudo-capacitive behavior. Accordingly, GCD curves at different specific currents in Fig. 5b show similar characteristics without obvious voltage plateau region. As shown in Fig. S11, the maximum specific capacitance value is calculated as 141 F g$^{-1}$ at 0.5 A g$^{-1}$ based on the total mass of NiAl-LDHs and YP-80F. It is noteworthy that the specific capacitance still remains 78 F g$^{-1}$ at 20 A g$^{-1}$ with a capacitance retention of 55.3%.

Fig. 5c shows the Ragone plot of the NAC-Ov//YP-80F HSC. The maximum specific energy is achieved as 50.0 W h kg$^{-1}$ at 400 W kg$^{-1}$. Meanwhile, as the specific power increases to 16,000 W kg$^{-1}$, the specific energy is retained as 27.2 W h kg$^{-1}$. These values are higher than or comparable to those of the NiAl-LDHs based HSCs in relevant literature, such as CC@NiCoAl-LDHs//ZPC HSC (44.0 W h kg$^{-1}$ at 462 W kg$^{-1}$) [35], NiAl-LDHs//ACNF HSC (20.0 W h kg$^{-1}$ at 750 W kg$^{-1}$) [36], CNP-LDHs//p-GN HSC (50.0 W h kg$^{-1}$ at 467 W kg$^{-1}$) [37], NiAl-LDO/LDS//G HSC (47.9 W h kg$^{-1}$ at 750 W kg$^{-1}$) [38], H-NiAl-LDHs//G (34.1 W h kg$^{-1}$ at 700 W kg$^{-1}$) [39], NA-LDH-OA//AC HSC (40.3 W h kg$^{-1}$ at 943 W kg$^{-1}$) [40], and NiCoAl-LDHNs//AC HSC (35.9 W h kg$^{-1}$ at 226 W kg$^{-1}$) [41]. Furthermore, the HSC exhibits excellent cyclability with a capacitance retention of 91% after 10,000 cycles (Fig. 5d).

### UOR activity of NiAl-LDHs/CFP electrodes

The electrocatalytic behavior of NiAl-LDHs/CFP electrodes for UOR was evaluated by linear sweep voltammetry (LSV) and chronopotentiometry (CP) measurements in a standard three-electrode system. LSV polarization curves of NAC-Ov for OER and UOR are compared in Fig. S12. An obvious peak at 1.49 V vs. RHE is observed for the OER curve due to the oxidation process of Ni$^{2+}$ state [42]. After adding urea (0.33 mol L$^{-1}$), the onset potential of NAC-Ov decreases significantly, indicating its superior activity for UOR. In addition, the oxidation peak disappears because of the overlap of large UOR current [43]. Fig. 6a shows the comparison of UOR electrocatalytic performance of NAC-U and NAC-Ov in 1 mol L$^{-1}$ KOH with 0.33 mol L$^{-1}$ urea. To drive a current density of 10 mA cm$^{-2}$, potential values of 1.42 V and 1.45 V vs. RHE are required for NAC-Ov and NAC-U, respectively, denoting the higher electrocatalytic activity of NAC-Ov for UOR in alkaline solution. The excellent UOR activity for the NAC-Ov is also comparable to that of recently reported nickel-based UOR electrocatalysts (Fig. 6b and Table S6).

Tafel plots were used to further assess the UOR kinetics of these NiAl-LDHs. As displayed in Fig. 6c, the calculated Tafel

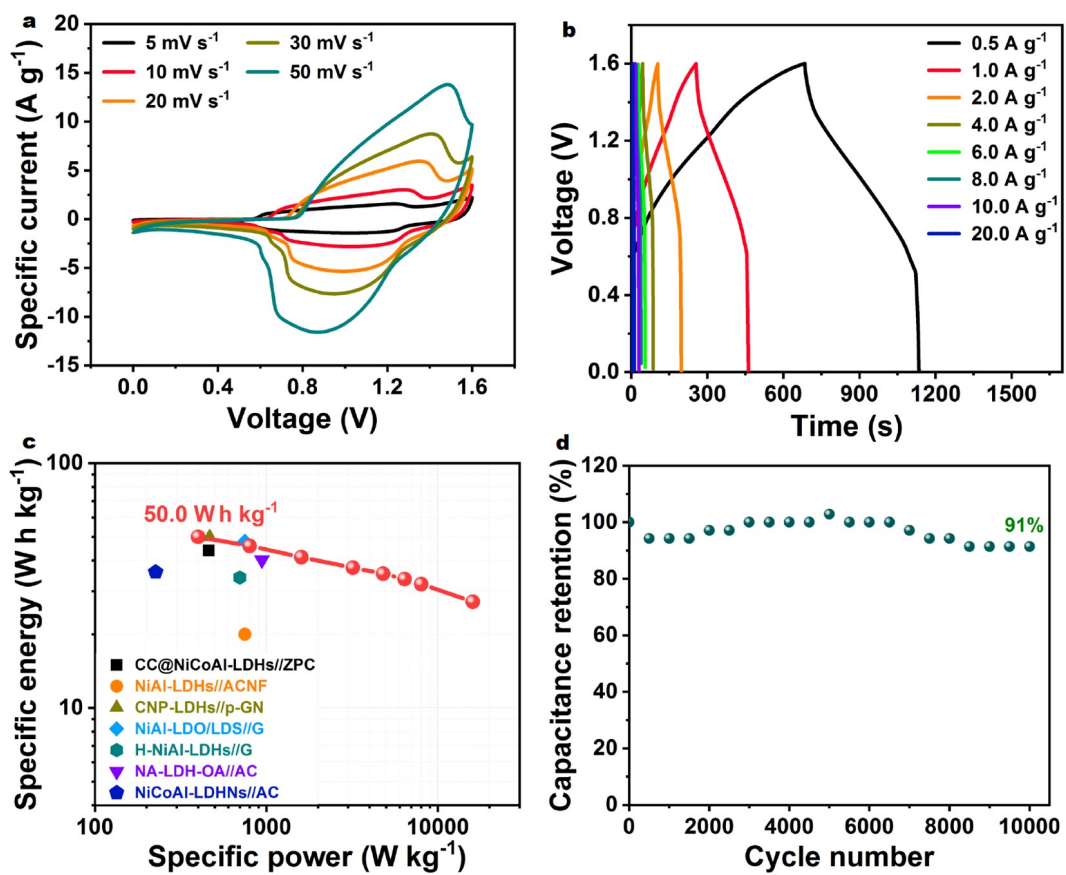

**Figure 5** (a) CV curves of the HSC at different scan rates from 5 to 50 mV s$^{-1}$. (b) GCD curves of the HSC at different specific currents from 0.5 to 20.0 A g$^{-1}$. (c) Ragone plot related to the specific energy and specific power of the HSC. (d) Cycle stability of the HSC at 10 A g$^{-1}$.

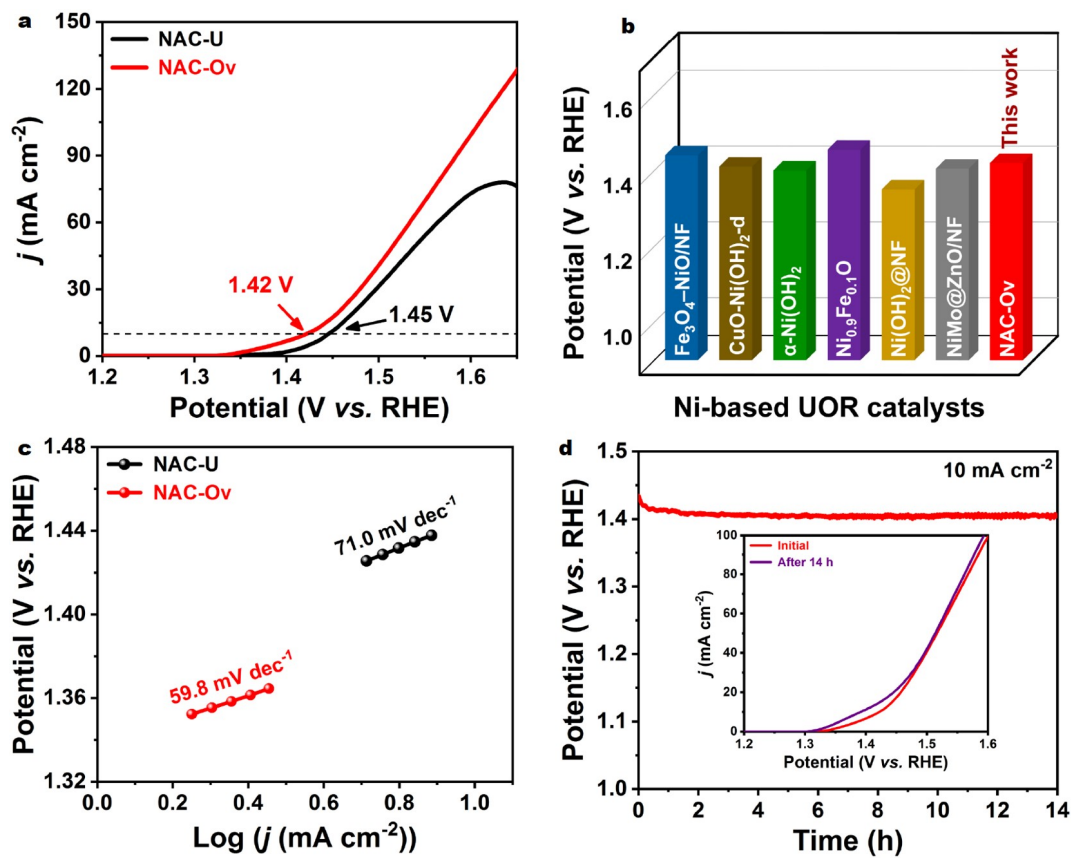

**Figure 6** (a) LSV polarization curves of NAC-U and NAC-Ov in 1 mol L$^{-1}$ KOH with 0.33 mol L$^{-1}$ urea at a scan rate of 5 mV s$^{-1}$. (b) Comparison of the potential value of NAC-Ov with other nickel-based UOR electrocatalysts at a current density of 10 mA cm$^{-2}$. (c) Corresponding Tafel plots. (d) CP curve of NAC-Ov at a constant current density of 10 mA cm$^{-2}$ for 14 h (inset: LSV polarization curves for NAC-Ov before and after CP measurements).

slope of NAC-Ov is 59.8 mV dec$^{-1}$, which is lower than that of NAC-U (71.0 mV dec$^{-1}$). The improved UOR kinetics of NAC-Ov was also verified by EIS results (Fig. S13). According to Table S7, the $R_{ESR}$ and $R_{ct}$ of NAC-Ov are lower than those of NAC-U, which is consistent with the EIS results in the supercapacitive analysis. Furthermore, the durability of NAC-Ov was evaluated via the CP test at a constant current density of 10 mA cm$^{-2}$. From Fig. 6d, a negligible potential increase is realized after 14 h. Subsequently, the LSV polarization curve is almost unchanged after 14 h as compared with the initial state, demonstrating the excellent electrochemical stability of NAC-Ov for UOR.

Based on the supercapacitive and electrocatalytic performance of NAC-U and NAC-Ov, the bifunctional NAC-Ov exhibits excellent charge storage ability and UOR activity. The outstanding electrochemical performance is attributed to the combined contribution of a large specific surface area and high oxygen-vacancy concentration. Specifically, the large exposed surface area of NAC-Ov can provide abundant active centers for charge storage and urea oxidation. Meanwhile, the existence of ample oxygen vacancies not only increases the electronic conductivity of the electrode ($R_{ESR}$), but also facilitates the diffusion of active species into the mesopores of the electrode ($R_{ct}$). Consequently, NAC-Ov can achieve an ultra-high specific capacitance of 2801 F g$^{-1}$ (700 C g$^{-1}$) at 1 A g$^{-1}$ for HSC and exhibit a low potential of 1.42 V vs. RHE at 10 mA cm$^{-2}$ for UOR.

## CONCLUSIONS

In summary, NiAl-LDH nanosheets with abundant oxygen vacancies were successfully synthesized and in situ grown on the surface of CFP without the use of any exfoliating solvents and binder. The NiAl-LDHs/CFP electrode displays superior supercapacitive performance with an ultra-high specific capacitance of 2801 F g$^{-1}$ at 1 A g$^{-1}$ and a desired stability of 80% capacitance retention after 4000 cycles and electrocatalytic performance towards UOR (a low potential of 1.42 V vs. RHE at 10 mA cm$^{-2}$, a Tafel slope of 59.8 mV dec$^{-1}$, and a robust durability over 14 h). Compared with the NiAl-LDHs prepared by the conventional hydrothermal method, the enhanced electrochemical performance of oxygen-deficient NiAl-LDH nanosheets is mainly attributed to the larger specific surface area (216.6 m$^2$ g$^{-1}$) and higher electronic conductivity (3.45 × 10$^{-4}$ S cm$^{-1}$). Therefore, this work provides the rational design of bifunctional energy materials for HSC and urea oxidation through tuning the oxygen-vacancy concentration of nanostructured nickel-based LDHs.

**Acknowledgements** This work was supported by the National Natural Science Foundation of China (21776324 and 22078374), Guangdong Basic and Applied Basic Research Foundation (2019B1515120058 and 2020A1515011149), National Ten Thousand Talent Plan, National Key R&D Program of China (2018YFD0800703 and 2020YFC1807600), Key-Area Research and Development Program of Guangdong Province

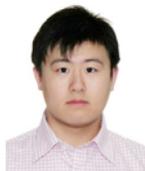

**Yuchen Wang** received his PhD degree in mechanical engineering from the University of Miami in 2017. He is currently a postdoc research associate at the School of Environmental Science and Engineering, Sun Yat-sen University. His present research interests focus on nanomaterials in energy storage and conversion applications.

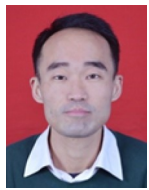

**Kai Yan** is a full professor at the School of Environmental Science and Engineering, Sun Yat-sen University. He received his PhD degree from Max-Planck-Institute for Coal Research and RWTH Aachen University in 2011. Then he obtained an Ontario Government Postdoctoral Fellowship at Lakehead University (2012–2013) and joined Brown University as a senior postdoctoral research associate (2013–2016). His current interests include the synthesis of nanostructured materials for clean energy and environment-related applications.

## 一步法构筑双功能富含氧空位的花瓣状镍铝水滑石纳米薄片用于高性能超级电容器和尿素氧化

王宇辰, 刘瑶钰, 张曼, 刘碧莹, 赵志月, 严凯[*]

**摘要**　镍基水滑石在能量存储(超级电容器)和转化(尿素氧化)领域是很有前景的电极材料. 合理构建镍基水滑石的原子和电子结构对于实现其理想的电化学性能至关重要. 本论文报道了一种简单、环境友好的一步法制备富含氧空位的花瓣状镍铝水滑石(NiAl-LDHs)纳米薄片用于混合超级电容器和尿素氧化. 性能最好的富氧空位NiAl-LDHs纳米薄片具有216.6 m$^2$ g$^{-1}$的大比表面积和3.45 × 10$^{-4}$ S cm$^{-1}$的高电导率, 可在1 A g$^{-1}$的比电流下展示出2801 F g$^{-1}$(700 C g$^{-1}$)的超高比电容. 基于NiAl-LDHs//商业活性炭组装的混合超级电容器在400 W kg$^{-1}$比功率密度下可获得50.0 W h kg$^{-1}$的比能量, 且循环10,000次后仍有91%的电容保持率. 同时, NiAl-LDHs纳米薄片作为高效的尿素氧化电催化剂, 在1 mol L$^{-1}$ KOH和0.33 mol L$^{-1}$尿素中仅需1.42 V *vs.*可逆氢电极的氧化电位就可达到10 mA cm$^{-2}$的电流密度. 由于NiAl-LDHs的纳米薄片结构暴露了更多的活性位点和丰富的氧空位, 其电化学性能优于大部分报道的镍基水滑石. 因此, 本研究为多功能纳米能源材料的合理设计奠定了良好的基础.